\documentclass[
  aps,
  prb,
  superscriptaddress,
  floatfix,
  twocolumn,
  secnumarabic,
  amssymb,
  amsmath]{revtex4-1}

\usepackage{dcolumn}
\usepackage{bm}
\usepackage{hyperref}

\usepackage[pdftex]{graphicx}
\usepackage{tabularx}
\usepackage{subfigure}
\usepackage{amssymb}
\usepackage{enumerate}
\usepackage{array}
\newlength{\gnupicwidth}
\usepackage{ifthen}
\usepackage{color}
\usepackage{ulem}             
\definecolor{rot}{rgb}{1,0,0}
\definecolor{blau}{rgb}{0,0,1}
\definecolor{orange}{rgb}{.5,.5,0}
\definecolor{dunkelgruen}{rgb}{.133,0.545,0.133}

\newif\ifcom
\newif\ifdel
\comtrue               %
\delfalse

\begin{document}

\title{Impact of the interface quality of Pt/YIG(111) hybrids on their spin Hall magnetoresistance}

\author{Sabine~P{\"u}tter}
\email[]{s.puetter@fz-juelich.de}
\affiliation{J{\"u}lich Centre for Neutron Science (JCNS) at Heinz Maier-Leibnitz  Zentrum (MLZ), Forschungszentrum J{\"u}lich GmbH, Lichtenbergstr.\ 1, 85747 Garching, Germany}
\author{Stephan~Gepr\"{a}gs}
\email[]{stephan.gepraegs@wmi.badw.de}
\affiliation{Walther-Mei{\ss}ner-Institut, Bayerische Akademie der Wissenschaften, 85748 Garching, Germany}
\author{Richard~Schlitz}
\affiliation{Walther-Mei{\ss}ner-Institut, Bayerische Akademie der Wissenschaften, 85748 Garching, Germany}
\affiliation{Physik-Department, Technische Universit\"{a}t M\"{u}nchen, 85748 Garching, Germany}
\author{Matthias~Althammer}
\affiliation{Walther-Mei{\ss}ner-Institut, Bayerische Akademie der Wissenschaften, 85748 Garching, Germany}
\author{Andreas~Erb}
\affiliation{Walther-Mei{\ss}ner-Institut, Bayerische Akademie der Wissenschaften, 85748 Garching, Germany}
\author{Rudolf~Gross}
\affiliation{Walther-Mei{\ss}ner-Institut, Bayerische Akademie der Wissenschaften, 85748 Garching, Germany}
\affiliation{Physik-Department, Technische Universit\"{a}t M\"{u}nchen, 85748 Garching, Germany}
\author{Sebastian~T.B.~Goennenwein}
\affiliation{Walther-Mei{\ss}ner-Institut, Bayerische Akademie der Wissenschaften, 85748 Garching, Germany}
\affiliation{Institut f\"{u}r Festk{\"o}rperphysik, Technische Universit\"{a}t Dresden, 01062 Dresden, Germany}

\date{\today}
\begin{abstract}
We study the influence of the interface quality of Pt/Y$_3$Fe$_5$O$_{12}$(111) hybrids on their spin Hall magnetoresistance. This is achieved by exposing Y$_3$Fe$_5$O$_{12}$(111) single crystal substrates to different well-defined surface treatments prior to the Pt deposition. The quality of the Y$_3$Fe$_5$O$_{12}$(YIG) surface, the Pt/YIG interface and the Pt layer is monitored \textit{in-situ} by reflection high-energy electron diffraction and Auger electron spectroscopy as well as \textit{ex-situ} by atomic force microscopy and x-ray diffraction. To identify the impact of the different surface treatments on the spin Hall magnetoresistance, angle-dependent magnetoresistance measurements are carried out at room temperature. The largest spin Hall magnetoresistance is found in Pt/YIG fabricated by a two-step surface treatment consisting of a ``piranha'' etch process followed by an annealing step  at $500^\circ$C in pure oxygen atmosphere. Our data suggest that the small SMR in Pt/YIG  without any surface treatments of the YIG substrate prior to Pt deposition is caused by a considerable carbon agglomeration at the Y$_3$Fe$_5$O$_{12}$ surface.  
\end{abstract}
\maketitle

In the field of spintronics, the efficient generation and detection of spin currents is fundamental for new memory and logic devices. Therefore, over the past years spin current transport has been extensively studied in paramagnetic (normal) metal (NM)/ferromagnetic insulator (FMI) hybrid structures in spin pumping,  spin Seebeck effect, or  spin Hall magnetoresistance (SMR) experiments.\cite{Urban2001,Hein11,Czeschka2011,Uchida2010,Gepraegs2016,Naka13,Alth13} In all these experiments the signal amplitude sensitively depends on the transfer of a spin current, i.e.\ spin angular momentum, across the NM/FMI interface and its interconversion into an electrical signal via the inverse spin Hall effect.\cite{Hirsch1999,Chen13b} 

According to theory, the relevant interface property determining the spin current flow across the NM/FMI interface is the spin mixing conductance.\cite{Brataas2000,Weiler2013} In several experiments it has been shown that the spin mixing conductance sensitively depends on the quality of the NM/FMI interface.\cite{Burr12,Jung13,Qiu13,Qiu15} For example, Jungfleisch \textit{et al.} reported an increase of the spin mixing conductance by more than two orders of magnitude using a combination of piranha wet etching and an \textit{in-situ} O$^+$/Ar$^+$ plasma treatment of the FMI surface prior to the NM deposition.\cite{Jung13} A clean and well-controlled NM/FMI interface can be obtained by \textit{in-situ} deposition of the NM layer subsequent to the FMI thin film growth without breaking the vacuum.\cite{Alth13} However, this procedure is often not possible if single crystal samples are used, which are superior to epitaxial thin films regarding structural and magnetic quality. In this case the NM layer is deposited \textit{ex-situ} on the single crystal, which is exposed to ambient conditions prior to the deposition resulting in adsorbed molecules, mainly carbon, on the surface. As a consequence the molecules may form additional spin-scattering centers and finally provoke a loss of spin information at the NM/FMI interface.

In this work we systematically investigate how different surface treatments of yttrium iron garnet (Y$_3$Fe$_5$O$_{12}$, YIG) single crystals prior to the Pt deposition impact the SMR  in Pt/YIG hybrid structures. Up to now, only indirect information on the  quality of the NM/YIG interface has been derived by e.g.\ measuring  the inverse spin Hall effect voltage in the NM layer.\cite{Jung13} A systematic investigation of the surface properties is still lacking. In our study we employ both \textit{in-situ} and \textit{ex-situ} surface and structural characterization methods to obtain reliable information of the influence of different surface preparation procedures on the surface viz.\ interface properties. We then correlate the observed SMR magnitude with the interface properties.

The YIG single crystals were grown using the traveling solvent floating zone (TSFZ) method in a 4-mirror image furnace.\cite{Lambacher2010} As a solvent in the crystal growth process a composition of about 20\,mol per cent of Y$_2$O$_3$ in YFeO$_3$ was used. Due to the high solubility of YIG in its solvent, the growth speed was as high as 4\,mm per hour. Single crystals of YIG with a diameter of about 5\,mm and a length of about 50\,mm were obtained. The crystals were cut into pieces with a diameter of about 5\,mm and a thickness of 1\,mm. These crystals were polished along the (111)-plane and used as a substrate for the deposition of thin Pt layers. The Pt deposition was carried out at room temperature by electron beam evaporation in a DCA M600 MBE system with a base pressure of $10^{-10}$\,mbar using a growth rate of around 0.3\,\AA/s. Prior to the deposition different surface treatments of the YIG substrates were carried out:
\begin{enumerate}[Procedure A:]
\item Cleaning in ethanol and isopropanol (denoted as ``raw'' YIG crystal)
\item Cleaning in ``piranha'' etch for 10\,min (denoted as ``etched'' YIG crystal)\cite{Piranha}
\item Additional annealing of the ``raw'' YIG crystal 
\item Additional annealing of the ``etched'' YIG crystal 
\end{enumerate}
The annealing was performed \textit{in-situ} in the MBE system at 500$^\circ$C for 40\,min in a pure oxygen atmosphere  of $p~=~10^{-5}$\,mbar. 
\begin{figure}
\includegraphics[width=1.0\columnwidth]{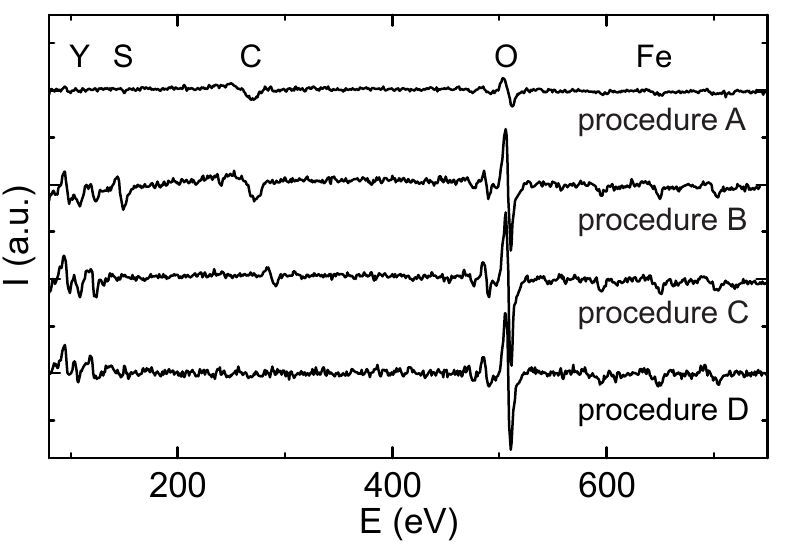}
 \caption{Auger electron spectra of YIG(111) single crystals carried out after performing different surface procedures: cleaning in ethanol and isopropanol (A), cleaning in ``piranha'' etch (B), procedure A with additional annealing (C) and procedure B with additional annealing (D). A polynomial background was subtracted from the raw data and the obtained curves are vertically shifted for clarity.
}
\label{fig:Puetter_Fig1}
\end{figure}

\newcolumntype{L}[1]{>{\raggedleft\let\newline\\\arraybackslash\hspace{0pt}}m{#1}}
\begin{table}[b]
\begin{center}
\begin{ruledtabular}
\begin{tabular}{*{1}{p{.02\linewidth}}
>{\centering\let\newline\\\arraybackslash}p{.05\linewidth}
>{\centering\let\newline\\\arraybackslash}p{.065\linewidth}
*{5}{>{\raggedleft\let\newline\\\arraybackslash}p{.05\linewidth}}}
	proc.&	etch.	       & anneal.      &	Y (\%)	&	O (\%) & Fe (\%) & C (\%)	& S (\%)	\\ \hline
  A	   & $-$	         & $-$	        & 7      &	22	   & 5	     & 66     &	0       \\ \hline
  B	   & $\checkmark$  & $-$	        & 23     &	29     & 9       & 36     &	3	      \\ \hline
  C    & $-$	         & $\checkmark$ & 24	    & 40     & 11      & 25     & 0	      \\ \hline
  D    & $\checkmark$	 & $\checkmark$	& 33      & 52     & 15      & 0      & 0       \\ 
\end{tabular}
\end{ruledtabular}
\end{center}
\caption{Elemental surface concentrations of YIG(111) single crystals obtained from AES using different surface treatments. The uncertainty of the elemental surface concentrations is estimated to about 5\%.}
\label{tab:YIG} 
\end{table}

 After the different cleaning procedures, the elemental surface
concentrations were determined by Auger electron spectroscopy
(AES) using an incident electron energy of 3\,keV. The obtained AES
spectra are shown in Fig.~\ref{fig:Puetter_Fig1}. The evaluation of
the data was carried out using the peak-to-peak Auger
amplitudes.\cite{Palm76,Powe90} The thus obtained elemental
concentrations are summarized in Table~\ref{tab:YIG}. As obvious from
Fig.~\ref{fig:Puetter_Fig1} and Table~\ref{tab:YIG}, the elemental
surface concentrations strongly depend on the surface treatment. While
carbon and oxygen dominate the surface of ``raw'' YIG crystals
(procedure A), the carbon concentration can be reduced and the yttrium and
iron concentrations can be increased by either using a piranha etch (procedure
B) or annealing the crystal in oxygen (procedure C). However, after
using procedure B we additionally detected a small amount of sulfur
caused by the piranha etch, which can be removed by a subsequent
annealing step (procedure D). Furthermore, the formation of carbide on  the
YIG surface indicated by the different shape and position of the carbon
peak was found after annealing the "raw" YIG crystal in oxygen (procedure
C).\cite{Davi78}

In fact, Fig.~\ref{fig:Puetter_Fig1} and
Table~\ref{tab:YIG} reveal that procedure D yields the purest YIG
surface, without carbon or sulfur contamination. However, the
elemental surface concentrations do not agree with the bulk concentrations
of yttrium (15\%), oxygen (60\%), and iron (25\%). In contrast, we
find 33\% of yttrium, 52\% of oxygen, and 15\% of iron. The deviation
might be explained by the different concentration of yttrium and iron
at the (111)-surface of YIG. Note that in YIG thin films fabricated by pulsed laser deposition (PLD) a Fe deficiency has also been observed.\cite{Hein11}

\begin{figure}[t]
\includegraphics[width=1.0\columnwidth]{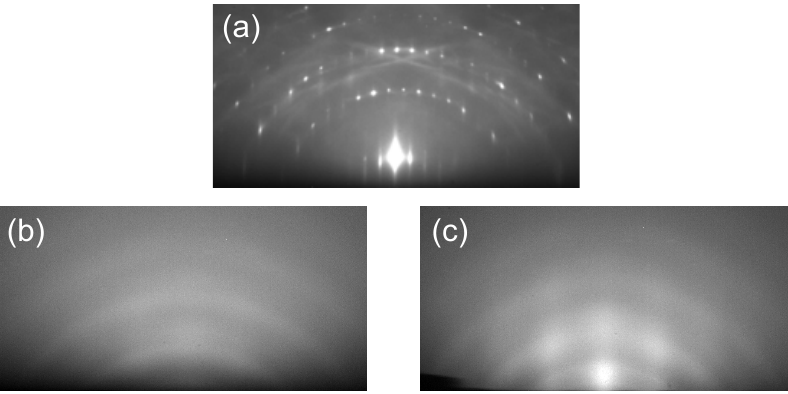}
 \caption{RHEED pattern of (a) a YIG(111) surface recorded after using surface treatment~D. (b),(c) RHEED patterns of 6 nm  Pt  after the deposition on YIG crystals using surface treatment~C and treatment~D prior to the deposition, respectively.}
\label{fig:Puetter_Fig2}
\end{figure}

Additional structural information of the surface was obtained by using \textit{in-situ} reflection high energy electron diffraction (RHEED) as well as low energy electron diffraction (LEED). While for samples with procedure~A neither RHEED nor LEED patterns were detected, for those with procedure~B a RHEED but no LEED pattern was obtained. Note that the absence of a RHEED and LEED pattern means that there is neither crystalline nor polycrystalline order within the respective probing depth of RHEED (about 10 \AA\,for 15 keV electrons at low angle of incidence) and LEED (about 5 \AA\,for 100\,-\,500\,eV). In contrast, for procedure~C and D LEED and RHEED patterns of similar quality were visible. Actually, the observation of well-defined spots provides evidence for low surface roughness and high crystallinity of the YIG(111) surface, cf.\ Fig.~\ref{fig:Puetter_Fig2}(a). This is corroborated by \textit{ex-situ} atomic force microscopy (AFM) experiments, yielding a surface roughness of only 1.6\,\AA\,(root mean square value) for this sample. In total, a carbon-free YIG surface with low roughness and high crystallinity can be obtained following procedure~D.

Subsequent to the different YIG surface treatments, about 6 nm thick Pt films were  deposited \textit{in-situ}, i.~e.\ without breaking  vacuum on the YIG crystals. The Pt deposition was monitored by RHEED. While, again, no RHEED pattern was obtained for the Pt thin films deposited on YIG crystals using procedure~A or B, the RHEED patterns of Pt thin films on YIG crystals prepared by procedure~C and D reveal intensity rings, demonstrating a polycrystalline nature of the Pt thin films (cf.~Fig.~\ref{fig:Puetter_Fig2}(b),(c)). Furthermore, weak spots visible in Fig.~\ref{fig:Puetter_Fig2}(c) indicate weakly-textured Pt thin films on YIG crystals using surface treatment~D. This is also confirmed by x-ray diffraction measurements (not shown here). Using Scherrer's formula the average size of the Pt crystallites can be estimated to about 10\,nm taking only size effects into account.\cite{Langford1978}

After deposition, the elemental surface concentrations of the Pt films were investigated using \textit{in-situ} AES. The results are summarized in Table~\ref{tab:YIGPt}. The information depth of the given elements is about 1.2\,nm. While pure Pt thin films were obtained on YIG crystals prepared by procedure~C or D, a carbon contamination is found in the Pt films deposited on YIG crystals using cleaning procedure~A or B. Obviously, the carbide formed after procedure~C remains at the YIG surface while the carbon seems to diffuse into the Pt film.
\begin{table}[t]
\begin{center}
\begin{ruledtabular}
\begin{tabular}{lrrrccc}
proc.	& Pt 	  &	O    &	C 	&	thickness  &$\rho_0$	  &	SMR       	      \\ 
      &  (\%) & (\%) & (\%) & (\AA)      & (n$\Omega$m) & ($10^{-4}$)       \\ \hline
A	    &	76    &	0	   &	24	&	63 $\pm$ 2 &	589      &	0.14	$\pm$	0.07	\\ \hline
B	    &	79	  &	0    &	21  &	61 $\pm$ 2 &	393	    &	1.88	$\pm$	0.10	\\ \hline
C	    &	100	  &	0	   &	0	  &	60 $\pm$ 2 &	408      &	1.24	$\pm$	0.02	\\ \hline
D	    &	100	  &	0	   &	0	  &	59 $\pm$ 1 &	353      &	3.48	$\pm$	0.01	\\ \hline
\end{tabular}\end{ruledtabular}
\end{center}
\caption{\label{tab:overview} Overview of the Pt/YIG(111) samples and their parameters. The elemental concentration of Pt, O, and C of the Pt thin film was obtained by AES and the thickness by X-ray reflectometry. The resistivity of the Pt layer $\rho_0$ and the SMR were determined by ADMR at 300\,K.}
\label{tab:YIGPt}
\end{table}

\begin{figure}
\includegraphics[width=1.0\columnwidth]{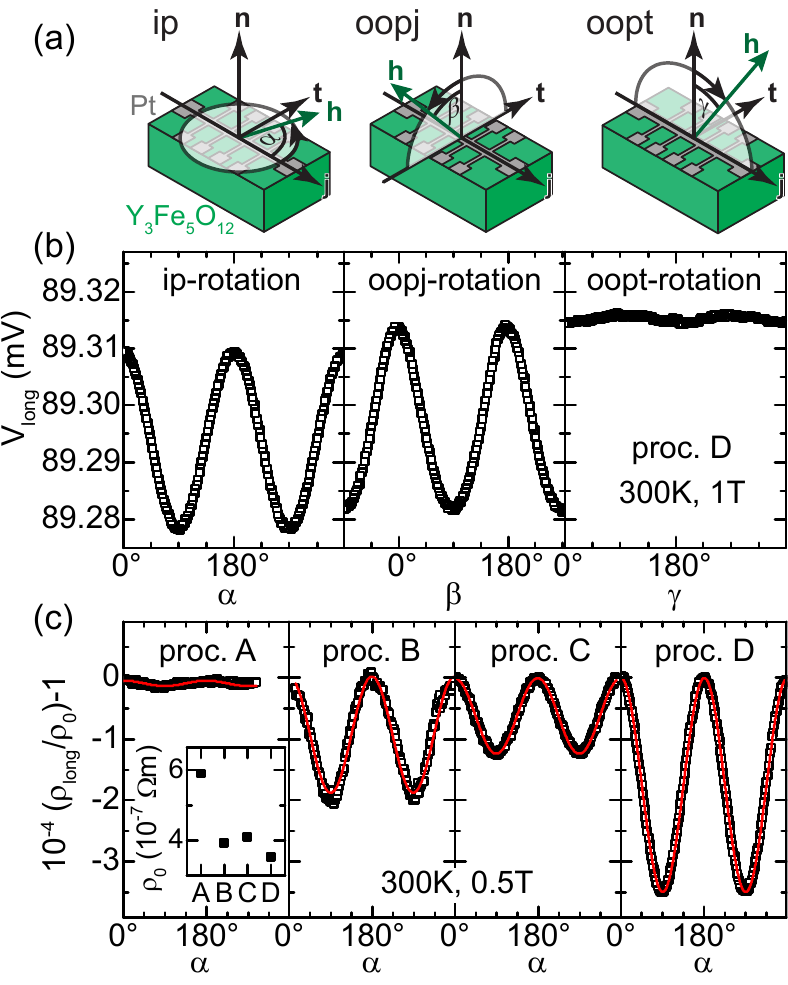}
 \caption{(Color online) ADMR measurements at 300\,K of Pt thin films deposited on YIG(111) crystals after different surface treatments. (a) Schematic of the Hall bar mesa structure, the coordinate system defined by $\mathbf{j}$, $\mathbf{t}$, and $\mathbf{n}$, as well as the different rotation planes of the magnetic field direction $\mathbf{h}=\mathbf{H}/\left|\mathbf{H}\right|$. (b) Angle-dependence of the longitudinal voltage $V_\mathrm{long}$ recorded on a Pt/YIG sample using YIG surface treatment D while rotating the magnetic field in the film plane (ip-rotation) and in the two orthogonal out-of-plane rotation planes (oopj- and oopt-rotations). Due to small temperature drifts, different maximum values of $V_\mathrm{long}$ were obtained in the ip-, oopj-, and oopt-rotation measurements. (c) SMR recorded while rotating the magnetic field in-plane of Pt/YIG samples prepared by using different surface treatments prior to the Pt deposition (procedure A-D). The red line represents a $\cos^2(\alpha)$-fit to the ADMR data  to extract the SMR magnitude. The inset shows the resistivity $\rho_0$ of the Pt layer as a function of different surface treatments of the YIG crystal (procedure A-D).}
 \label{fig:Puetter_Fig3}
\end{figure}

 To identify the impact of the different surface treatments on the
YIG substrates on the SMR effect, angle-dependent magnetoresistance
(ADMR) measurements were carried out.\cite{Alth13} To this end the Pt
films were patterned into Hall bar shaped mesa structures using
photolithography and argon-ion beam milling. The ADMR
measurements were carried out in a liquid-He magnet cryostat at
300\,K. The magnetoresistance of the Pt thin film was determined by
applying a constant dc current of $I=200\,\mu$A along the Hall bar and
recording the longitudinal voltage signals $V_\mathrm{long}$, while
rotating the magnetic field in the
film plane (ip-rotation) as well as in two orthogonal out-of-plane
rotation planes (oopj- and oopt-rotation) at constant external magnetic field magnitudes  of 500\,mT and
1000\,mT (cf.~Fig.~\ref{fig:Puetter_Fig3}(a)). These magnetic field values are  both well above the saturation field of YIG. The longitudinal resistivity
can then straightforwardly be calculated to $\rho_\mathrm{long} =
V_\mathrm{long} w d_\mathrm{Pt} /(I l)$ using the width 
($w=80\;\mu$m) and the length ($l=600\;\mu$m) of the Hall bar mesa
structure as well as the thickness $d_\mathrm{Pt}$ of the Pt layer
(cf.~table~\ref{tab:YIGPt}).

As an example, Figure~\ref{fig:Puetter_Fig3}(b) shows the angle-dependence of $V_\mathrm{long}$ recorded from a Pt/YIG sample prepared by surface treatment~D.
Clearly, an ADMR is observed for rotations of the magnetic field in-plane (ip-rotation) as well as out-of-plane perpendicular to the current density direction (oopj-rotation), while almost no ADMR can be detected on rotating the magnetic field out-of-plane parallel to the direction of the applied current  (oopt-rotation). This is the characteristic fingerprint of the SMR, which can be phenomenologically described by $\rho_\mathrm{long} = \rho_0 + \rho_1(1- m_t^2)$, with $m_t$ being the projection of the normalized YIG magnetization  $\mathbf{m}=\mathbf{M}/\left|M\right|$ on $\mathbf{t}$, see  Fig.~\ref{fig:Puetter_Fig3}(a) for illustration of the coordinate system.\cite{Chen13b} We use a $\cos^2(\alpha)$ fit to the ADMR data obtained at 500~mT to extract the $\rho_0$ and $\rho_1$ values. According to the theoretical SMR model, the SMR magnitude is then defined as $\rho_1/\rho_0$.\cite{Chen13b,Alth13} Since $\rho_\mathrm{long}$ of the conventional anisotropic magnetoresistance (AMR) depends on $m^2_j$ and not on $m^2_t$, the 
 finite angle dependence of  $\rho_\mathrm{long}$ in the oopj-rotation plane and the vanishing angle dependence in the oopt-rotation plane (cf.\ Fig.~\ref{fig:Puetter_Fig3}(b)) clearly indicate that the present angle-dependent magnetoresistance is based on the spin Hall magnetoresistance.\cite{Chen13b,Alth13}

As obvious from Fig.~\ref{fig:Puetter_Fig3}(c), the SMR value as well as the resistivity of the Pt thin film is strongly dependent on the YIG surface treatment and thus the quality of the Pt/YIG interface. Only a small SMR value of $(0.14 \pm 0.07) \cdot 10^{-4}$ as well as a high resistivity of $(589 \pm 1) \mathrm{n}\Omega\mathrm{m}$ is observed in Pt thin films fabricated on as-received YIG crystals (procedure~A). 
This can be attributed to the high carbon contamination found in the Pt thin films (see table~\ref{tab:YIGPt}) enhancing the formation of grain boundaries,
\cite{Hira02} which increases the thin film resistivity.\cite{Tela02,Batt05,Gao11} Furthermore, the finite carbon contamination might also reduce the spin diffusion length in the Pt thin film, which weakens the SMR effect.  The SMR magnitude can be significantly increased by etching or annealing the YIG crystals prior to Pt deposition (procedure~B and C).  While the
Pt thin film on the YIG substrate prepared by procedure C is chemically clean, the carbide found on the YIG surface prepared according to procedure~C might act as a spin current barrier at the Pt/YIG interface.  

However, the largest SMR value of $(3.48 \pm 0.01) \cdot 10^{-4}$ as well as the lowest resistivity of $(353 \pm 1) \mathrm{n}\Omega\mathrm{m}$ are obtained by using the YIG surface treatment~D prior to the Pt deposition. The SMR value is close to the respective SMR value of YIG/Pt thin film bilayers fabricated by \textit{in-situ} deposition of Pt.\cite{Alth13} %
Our results demonstrate that the best interface with the highest spin Hall magnetoresistance is obtained by using a two-step treatment of the YIG crystal: In the first step the piranha etch reduces the carbon contamination of the YIG surface. Subsequent annealing in oxygen atmosphere results in an increase of the Fe content as well as a vanishing carbon and sulfur content at the surface.                        
 
In summary, we experimentally investigated the SMR in Pt thin films on YIG single crystals using different surface treatments of the YIG crystal prior to the deposition of Pt. We found an almost vanishing SMR value in Pt/YIG samples without any surface treatment of the YIG crystal and attribute this to a significant carbon contamination of the YIG surface and in the Pt thin film. The SMR value can be significantly increased by cleaning the YIG crystal using a piranha etch or by annealing the YIG crystal in oxygen. However, in the former case, we found a contamination with sulfur, while in the latter  the formation of carbide on the YIG surface was detected. The highest SMR value, which is comparable to that of \textit{in-situ} grown Pt/YIG bilayers,\cite{Alth13} was found for samples using a combination of etching and annealing of the YIG crystal prior to the Pt deposition. Our work demonstrates the high relevance of the interface quality for spin current based experiments and provides instructions for improving the interface quality. We thus point the way how to improve future spin current based devices. 

We thank K. Helm-Knapp and A. Habel for technical support and greatefully acknowledge financial support by the Deutsche Forschungsgemeinschaft via SPP 1538 (project no. GO 944/4) and the German Excellence Initiative via the "Nanosystems Initiative Munich (NIM)".

\bibliography{lit-Pt-YIG}

\begin{thebibliography}{25}
\bibitem[{Pir()}]{Piranha}
\bibinfo{note}{For the piranha acid we used concentrated sulfuric acid
  (H$_2$SO$_4$, 98\%) and hydrogen peroxide solution (H$_2$O$_2$, 30\%) in the
  ratio 1:1, by volume.}
\end{thebibliography}

\end{document}